\documentclass[aps,prl,twocolumn,superscriptaddress,groupedaddress,amsmath,amssymb,nofootinbib,floatfix]{revtex4}  
\usepackage{graphicx}  % needed for figures
\usepackage{dcolumn}   % needed for some tables
\usepackage{bm,color}        % for math
\usepackage{amssymb}   % for math

\newcommand{\be}{\begin{equation}}
\newcommand{\ee}{\end{equation}}
\def\CL{CosmoLib$2^{\rm nd}$} 

\begin{document}

\title{The CMB bispectrum from recombination}
\author{Zhiqi Huang and Filippo Vernizzi}
\address{CEA, Institut de Physique Th{\'e}orique,
         F-91191 Gif-sur-Yvette c{\'e}dex,  
         CNRS, Unit{\'e} de recherche associ{\'e}e-2306, F-91191 Gif-sur-Yvette c{\'e}dex}
\date{\today}

%%%%%%%%%%%%%%%%%%%%%%%%%%%%%%%%%%%%%%%%%%%%%%%%%%%%%%%%%%%%%%%%%%%%%
%%%% Abstract

\begin{abstract}
We compute the cosmic microwave background temperature bispectrum generated by nonlinearities  at recombination   on all scales. We use CosmoLib$2^{\rm nd}$, a numerical   Boltzmann code at second-order to compute CMB bispectra on the full sky. We consistently  include all effects except gravitational lensing, which can be added to our result using standard methods. The bispectrum is peaked on squeezed triangles and agrees with the  analytic approximation  in the squeezed limit at the few per cent level for all the scales where this is applicable. On smaller scales, we recover previous results on perturbed recombination. For cosmic-variance limited data to $l_{\rm max} =2000$, its signal-to-noise is $S/N=0.47$ and will bias a local signal by $f_{\rm NL}^{\rm loc}\simeq 0.82$. 
\end{abstract}

\maketitle

%\section{\label{sec:level1}First-level heading}
% sections are not used for PRL papers

{\em \bf Introduction:} The non-Gaussianity of cosmological perturbations is a powerful discriminator of early universe models. In particular, a future detection of local-type non-Gaussianity in the forthcoming Planck data \cite{planck} will rule out all single-field models with an adiabatic attractor and Bunch-Davies vacuum. However, nonlinearities connecting the initial conditions to the observed CMB can potentially contaminate the primordial signal and must be fully understood. 
The most important of these effects are at late-time, such as integrated Sachs-Wolfe (ISW)-lensing correlation \cite{Hanson:2009kg,Lewis:2011fk}, but second-order sources during recombination may also be relevant. 

Computing the full bispectrum from recombination requires to resort to a numerical Boltzmann code. Although there has been an intense effort to derive the complete second-order equations \cite{Bartolo:2005kv,Pitrou:2008hy,Senatore:2008vi,Pitrou:2010sn,Beneke:2010eg}, getting the bispectrum is very complicated and complete control of all effects has not been reached yet. 
Currently, the only complete numerical calculation is from the code CMBquick \cite{cyril}, which claims a contamination of $f_{\rm NL}^{\rm loc} \sim 5$ from recombination  \cite{Pitrou:2010sn}. However, this is in tension with partial calculations focussing on particular limits. For super-horizon modes at recombination the contamination is comparable to $f_{\rm NL}^{\rm loc} \sim -1/6$ \cite{Bartolo:2004ty,Boubekeur:2009uk}. On sub-Hubble scales two effects have been computed: fluctuations in the free-electron density, inducing local-type non-Gaussianity by delaying the time of recombination \cite{Khatri:2008kb,Senatore:2008wk}, and the nonlinear evolution of dark matter inducing equilateral-type non-Gaussianity in the photon temperature fluctuations \cite{Pitrou:2008ak,Bartolo:2008sg}. None of these effects will relevantly bias a local signal detection. 

Given the current status and the relevance of the result, it is important to shed some clarity on the full bispectrum from recombination, estimate its
contamination  and detectability. 
In this letter we compute the  full-sky bispectrum from recombination on all scales. We use \CL, a second-order numerical Boltzmann code which will be made publicly available in the future.

Fortunately, we have a robust theoretical guidance which allows to put the code on solid grounds. The bispectrum generated at recombination can be computed analytically in the squeezed limit and when one of the three modes is outside the horizon at recombination \cite{Creminelli:2004pv,Creminelli:2011sq,Bartolo:2011wb,Lewis:2012tc}. The argument is analogous to that of the single-field consistency relation \cite{Maldacena:2002vr}. 
After inflation, the effect of a super-horizon mode can be traded with a scale redefinition on the short ones. Thus, for scale-invariant perturbations three modes of the primordial curvature perturbation $\zeta$ are decorrelated in the squeezed limit. Of course, this is true only if $\zeta$ appears exponentially in the spatial metric. Analogously, since $T\propto 1/a$ the bispectrum of three super-horizon modes of $\tilde \Theta$ defined by $T = \bar T e^{\tilde \Theta}$ vanishes in the squeezed limit. The nonlinear relation converting the ``Gaussian'' variable $\tilde \Theta$ to the usual temperature fluctuation
$\Theta \equiv \delta T/ \bar T$ gives a local modulation inducing a temperature reduced bispectrum 
\be
\label{loc_red}
b_{l_1 l_2 l_3} =  C_{l_1} C_{l_2} + C_{l_1} C_{l_3} +C_{l_2} C_{l_3} + \tilde b_{l_1 l_2 l_3} \;,
\ee
where $C_{l}$ is the temperature spectrum and $\tilde b_{l_1 l_2 l_3}$ is the reduced bispectrum of $\tilde \Theta$.
Moreover, local physics during recombination is unaffected by the presence of a super-horizon mode of $\zeta$, which just acts as an unobservable scale redefinition. However, when the long mode eventually re-enters the horizon, it  induces an angular rescaling of the observed 2-point function in different regions in the sky. 
As the temperature spectrum substantially departs from scale-invariance due to oscillations in the photon-baryon plasma,
this induces a temperature bispectrum for $\tilde \Theta$ given by 
\be
\label{rescaling}
\tilde b_{l_1 l_2 l_3} = - \frac12 C^{\zeta T }_{l_1} \bigg( C_{l_2} \frac{d \ln l_2^2  C_{l_2}}{d \ln l_2} + C_{l_3} \frac{d \ln l_3^2  C_{l_3}}{d \ln l_3} \bigg) \;,
\ee
where $C^{\zeta T }_{l_1}$ is the cross-correlation spectrum between $\Theta$ and $\zeta$ at recombination and $ l_1 \ll l_{2,3}$. The symmetrization guarantees that corrections to this expression are of order  ${\cal O} ({l_1}/{l_{2,3}})^2$. Corrections due to sub-Hubble physics scale as ${\cal O} ({l_1}/{l_H})^2$, with $l_H = (Ha/c_s)_{\rm rec} \eta_0  \simeq 110$ \cite{Lewis:2012tc}.

The local redefinition \eqref{loc_red} and the spacial rescaling \eqref{rescaling} are not the only contributions to the bispectrum in the squeezed limit. Lensing close to the last scattering surface displaces the 2-point function inducing an isotropic and anisotropic rescaling of the coordinates \cite{Creminelli:2004pv,Boubekeur:2009uk,Lewis:2011fk}. 
Although in our numerical calculation we do not include lensing, the separation is clear: eqs.~\eqref{loc_red} and \eqref{rescaling} are the analytic approximation in the squeezed limit to the bispectrum computed by integrating the photon distribution along {\em unperturbed} geodesics.

\vskip.1cm
{\em \bf The code:} 
\CL~is a Boltzmann code based on CosmoLib \cite{Huang:2012mt}. It solves Einstein and Boltzmann equations in Poisson gauge for all cosmological species up to second-order in the  perturbations and computes the full-sky CMB bispectrum by integrating the second-order photon brightness along the line of sight and projecting it into harmonic space.

For the Boltzmann solver  we have employed the equations given in Ref.~\cite{Pitrou:2010sn} complementing them with those derived in several other references \cite{Bartolo:2005kv,Pitrou:2008hy,Senatore:2008vi,Beneke:2010eg}, paying attention to correct typos. The initial conditions are set on super-horizon scales assuming a minimal model of inflation where the primordial curvature perturbation $\zeta$ is approximately Gaussian. More generic adiabatic initial conditions are possible. Perturbed recombination is included using DRECFAST \cite{Novosyadlyj:2006fw}. The perturbation in the free-electron density  perfectly agrees with that computed by \cite{Senatore:2008vi} on large and small scales; $\lesssim 10\%$ disagreement is registered on scales of  order Hubble at recombination, due to the inconsistent treatment of metric perturbations in DRECFAST. This will be corrected in the future.

Second-order solutions with one of the modes super-horizon computed by the code agree with those obtained by a coordinate transformation as explained in \cite{Creminelli:2011sq}. On small scales and matter dominance, they agree with the analytic solutions that are known in the literature  \cite{Boubekeur:2008kn}.  
Moreover, we have extensively used CMBquick to compare our Boltzmann solutions.
Improvements with respect to such a code are: more precise treatment of the tight-coupling approximation necessary to evolve the early-time solutions from the initial conditions; different line-of-sight integration which reduces the errors due to truncations in the multipole expansion of the source, as described below; bispectrum computed for the full sky and with a different integration method.

{\em \bf Bispectrum computation:} 
The photon  brightness is defined from the photon one-particle distribution $f$ as $I (\eta,\vec x, \hat n) \equiv \int dp \, p^3 f (\eta, \vec x, \vec p) $, where $p^i$ is the momentum of photons in the local inertial frame and $p^i \equiv p n^i$, $n_i n^i=1$.
We use the metric in Poisson gauge, $g_{00} = - a^2(\eta) e^{2 \Phi}  $, $g_{0i}= a^2(\eta) \omega_i$ and $g_{ij} =  a^2(\eta)(e^{-2 \Psi} \delta_{ij} + \chi_{ij})$,
with $\omega_{i,i}=0$ and $\chi_{ii}=0=\chi_{ij,j}$. From the Boltzmann equation one can write an equation for the fractional brightness $\Delta \equiv \delta I/\bar I$,
\be
\label{brightness_eq}
\frac{d}{d \eta} (\Delta + 4 \Phi)   
 - 4 \Delta (\dot \Psi - \Phi_{,i} n^i) -E 
 =   -   \dot \tau  (1+\delta_e + \Phi) F\;, 
\ee
where $d/d\eta$ is the advective derivative, a dot denotes the derivative with respect to $\eta$ and $\dot \tau$ is the {\em unperturbed} differential optical depth.
On the l.h.s., $E \equiv  4 (\dot \Phi + \dot \Psi) - 4 \dot \omega_i n^i - 2 \dot \chi_{ij} n^i n^j $ is the redshift in  photon's energy due to integrated effects: the ISW contribution (whose second-order part is the Rees-Sciama (RS) effect), and the  vector and tensor contributions, respectively \cite{Boubekeur:2009uk}.  
The r.h.s.~comes from the collision term of the Boltzmann equation: $F$ can be read off, for instance, from the r.h.s.~of eq.~(78) of \cite{Senatore:2008vi}, with the notation for $\Phi$ and $\Psi$ interchanged; $\delta_e$ is the perturbation of the free-electron density.

The second term on the l.h.s.~of \eqref{brightness_eq} is problematic: it must be integrated along the line of sight and couples all multipole moments of the fractional brightness to the gravitational potentials. A crucial simplification comes from trading this term for a boundary term and one proportional to $\dot \tau$, contributing only at recombination. By replacing $\Phi_{,i} n^i $ by $ d \Phi/d\eta - \dot \Phi$ and dividing eq.~\eqref{brightness_eq} by $1+\Delta$ we can rewrite it, up to second order, as
\be
\label{brightness_eq2}
\begin{split}
&e^{- \int_{\eta}^{\eta_0} d \eta' \dot \tau (1+ \delta_e + \Phi  ) } \frac{d}{d \eta} \left[ \left(\tilde \Delta + 4 \Phi \right)  e^{ \int_{\eta}^{\eta_0} d \eta' \dot \tau  (1+ \delta_e + \Phi  ) } \right] \\
& - E =    -   \dot \tau   R \equiv -\dot \tau (1+ \delta_e +\Phi) (\tilde F + \tilde \Delta + 4 \Phi ) \;.
\end{split}
\ee
We have introduced the variable $\tilde \Delta \equiv \Delta - \Delta^2/2$, defined from the brightness in exponential form, $I = \bar I e^{\tilde \Delta}$, and $\tilde F \equiv F (1-\Delta) $. As explained above, $\tilde \Delta$ is Gaussian in the squeezed limit and on super-horizon scales. 
Analogously, one can show that $\tilde F $ on the r.h.s.~of \eqref{brightness_eq2} becomes Gaussian in the same limit \cite{ZF}.
Integrating this equation formally yields
\begin{align}
\label{eq_source}
&\tilde \Delta (\eta_0, \vec x =0 , \hat n)=  \int_{0}^{\eta_0} d \eta e^{-\tau (\eta) }  S(\eta, \vec x(\eta), \hat n )\;,\\
\label{source}
& S   \equiv  \left( E - \dot \tau  R \right) \left( 1 + \int_{\eta}^{\eta_0} d \eta' \dot \tau  \left( \delta_e + \Phi  \right) \right)  \;,
\end{align}
where $S$ is computed along photon trajectories. Here we integrate along a straight trajectory $\vec x (\eta) =  \hat n(\eta - \eta_0)$. In doing this we are neglecting the  Shapiro time-delay and gravitational lensing effects. While the former can be safely neglected \cite{Hu:2001yq}, gravitational lensing can be taken into account using standard treatments \cite{Lewis:2011fk,Hanson:2009kg}. Consistent inclusion of  lensing and time-delay effects in the code is left for  future work. 

The observed brightness $\tilde \Delta (\eta_0,\hat n)$ in \eqref{eq_source} can be decomposed into spherical harmonics, 
\be
\tilde \Theta (\eta_0, \hat n)  = \frac14 \tilde \Delta (\eta_0, \hat n) = \sum_{lm} \tilde a_{lm}  Y_{lm}(\hat n) \;,
\ee
where the factor of $1/4$ in the first equality comes from the relation between temperature and brightness, $I \propto T^4$. Correlating three $\tilde a_{lm}$ defines the bispectrum of $\tilde \Theta$, $\tilde B^{m_1 m_2 m_3}_{l_1l_2l_3}= {\cal G}_{l_1 l_2 l_3}^{m_1 m_2 m_3}  \tilde b_{l_1 l_2 l_3}$, where  $\tilde b_{l_1 l_2 l_3} $ is its reduced bispectrum and ${\cal G}_{l_1 l_2 l_3}^{m_1 m_2 m_3}$ are Gaunt coefficients \cite{Komatsu:2001rj}. The nonlinear conversion from $\tilde b_{l_1l_2l_3}$ to $ b_{l_1l_2l_3}$, the usual reduced bispectrum for temperature fluctuations $\Theta=\delta T/\bar T$,  is given by eq.~\eqref{loc_red}.

\vskip.1cm
{\em \bf The squeezed limit:} Locally, the effect of a long wavelength mode of $\zeta$ can be seen as a redefinition of the coordinates for the short modes \cite{Maldacena:2002vr}. At recombination, the long mode shifts the function $e^{-\tau}$ in \eqref{eq_source} through a time redefinition, $e^{-\tau } \to e^{-\tau } + \epsilon  (e^{-\tau})^{\hbox{$\cdot$}}$, where $\epsilon$ is the redefinition of the time coordinate, $\eta \to \eta + \epsilon(\eta)$ \cite{Creminelli:2011sq}, induced by the long mode. Equivalently, this can be seen as a long-wavelength perturbation in the free-electron density \cite{Senatore:2008vi},
\be
\label{deltae_long}
\delta_e + \Phi = (\dot \tau \epsilon )^{\hbox{$\cdot$}} \;.
\ee
The coordinate redefinition acts also on the source $S$ in eq.~\eqref{source}, generating second-order nonlinearities. For instance, it acts on  $\tilde F$ as $\tilde F \to \tilde F+ \epsilon \dot {\tilde F} + \zeta x^i \partial_i \tilde F$,\footnote{In the limit where one of the mode is super-horizon, quadratic terms in the velocity $\vec v$ or products of $\vec v$ times $\Delta_{lm}$ with $l>0$ in $\tilde F$ and vector and tensor perturbations in $E$ become negligible.} where the second term comes from a rescaling of the spacial coordinates, $x^i \to x^i(1+ \zeta)$ \cite{Creminelli:2011sq}. While $\tilde \Delta$ and $ \Phi $ behave as $\tilde F$, $E \simeq 4(\dot \Phi + \dot \Psi)$ transforms as its time derivative, $E \to E+ (\epsilon  E)^{\hbox{$\cdot$}} + \zeta x^i \partial_i E $. Using these properties and eq.~\eqref{deltae_long} in eq.~\eqref{source}, one sees that the long mode simply induces a spacial coordinate rescaling of the source $S$, 
\be
S \to S + \zeta x^i \partial_i S \;,
\ee
while the time redefinition is re-absorbed by the long wavelength free-electron density \eqref{deltae_long}. The spacial coordinate redefinition above yields the bispectrum of eq.~\eqref{rescaling}.
\begin{figure}[t]
\begin{center}
{\includegraphics[scale=1]{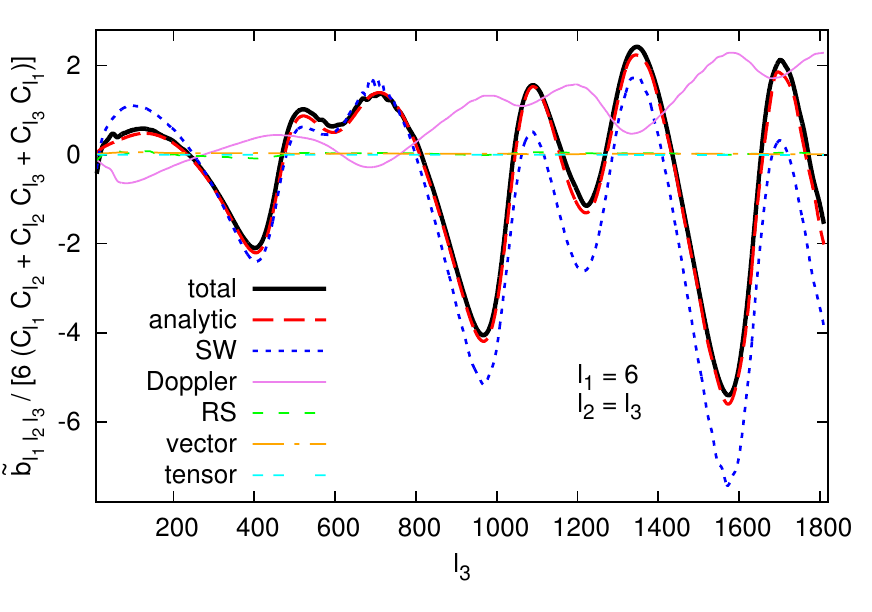}}
\caption{Reduced bispectrum of $\tilde \Theta$ for $l_1=6$ as a function of $l_2=l_3$  (black line),  compared to the analytic approximation from the spacial coordinate rescaling  \eqref{rescaling} (dashed-red line). The other lines are the different contributions to the bispectrum. In the squeezed limit, i.e.~for $l_3 \gg 6$, the agreement is at the few $\%$ level.}
\label{fig:1_l_fixed}
\end{center}
\end{figure} 
\begin{figure}[t]
\begin{center}
{\includegraphics[scale=1]{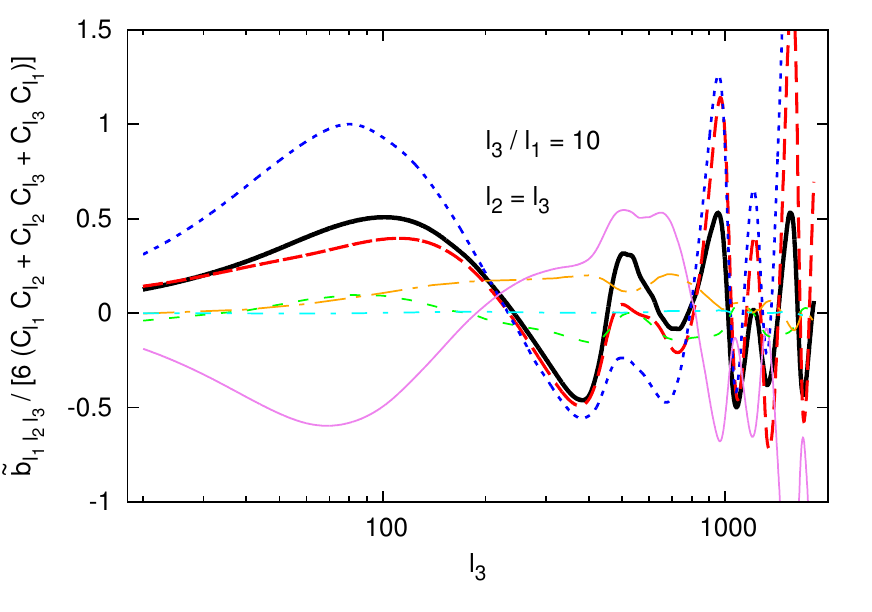}}
\caption{Reduced bispectrum for $l_1=l_3/10$ as a function of $l_2=l_3$. At small $l$ the bispectrum satisfies eq.~\eqref{rescaling}; deviations around $l_3\sim 100$ are due to integrated effects such as vector and RS. 
For $l_1 \gtrsim 50 \ (l_3 \gtrsim 500)$ the long mode is inside the Hubble radius at recombination and the analytic approximation fails.}
\label{fig:fixed_ratio}
\end{center}
\end{figure} 
\begin{figure}[t]
\begin{center}
{\includegraphics[scale=1]{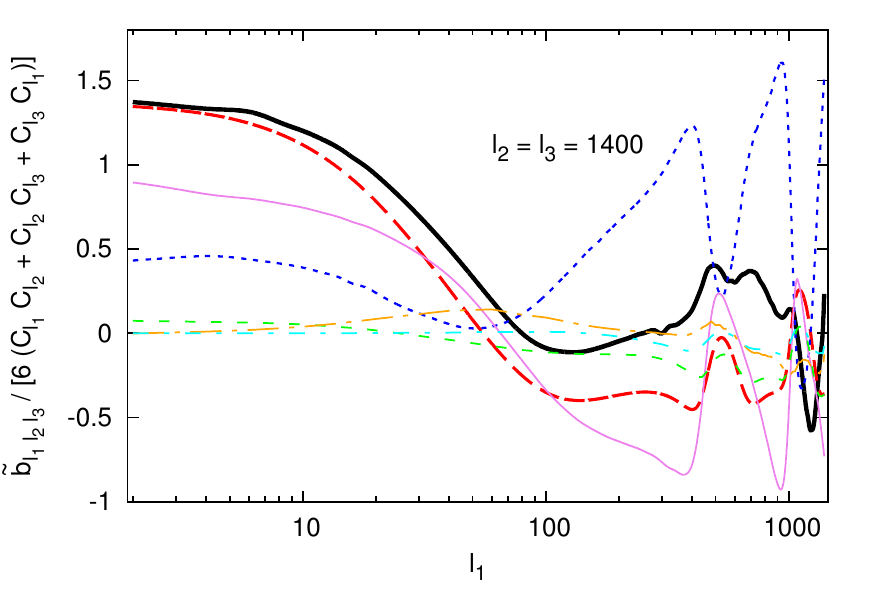}}
\caption{The reduced bispectrum with two $l$ fixed, $l_2=l_3$ as a function of $l_1$. As expected, for $l_1\lesssim 100$ the difference between the computed bispectrum and the analytic approximation is of order $\sim(l_1/l_H)^2$.}
\label{fig:fixed2l}
\end{center}
\end{figure} In Figs.~\ref{fig:1_l_fixed}, \ref{fig:fixed_ratio} and \ref{fig:fixed2l} we compare the reduced bispectrum of $\tilde \Theta$, $\tilde b_{l_1l_2l_3}$, computed using  \CL, with the analytic approximation \eqref{rescaling}. For this comparison we set $\Omega_{\Lambda}=0$ and  use the following cosmological parameters:  $\Omega_b=0.15$, $\Omega_c = 0.85$, $h=0.6$, $\tau_{\rm re} = 0.0$, $A_s = 2.4 \cdot 10^{-9}$ and $n_s=0.97$. This ensures that second-order late-time integrated effects do not correlate with late ISW, which cannot be taken into account by a coordinate redefinition at recombination.  We have split the bispectrum produced by the code into its different second-order contributions: SW, Doppler, Rees-Sciama, vectors and tensors. The bispectrum is dominated by the sum of the SW and Doppler effects, which cancel themselves due to acoustic oscillations.
Where applicable, the agreement between the computed bispectrum and the analytic approximation is remarkable. Residual few per cent deviations are mainly imputable to the inaccuracy of the 3-d integrator projecting the bispectrum in harmonic space.

\vskip.1cm
{\em \bf Shape and amplitude of the bispectrum:}
\begin{figure}[t]
\begin{center}
{\includegraphics[scale=0.7]{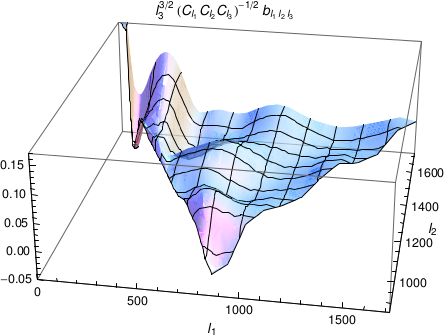}}
\caption{Signal-to-noise density, $l_3^{3/2} b_{l_1l_2l_3}/(C_{l_1} C_{l_2} C_{l_3})^{1/2}$, for $l_3 = 1720$ and $l_1\le l_2 \le l_3$. The signal is peaked on squeezed triangles and the acoustic oscillations are clearly displayed. For $l_1\ge 100$ the amplitude and shape agree with that computed in \cite{Senatore:2008wk}. }
\label{fig:3dplot}
\end{center}
\end{figure} 
We now numerically compute the bispectrum from recombination and compare it to a local bispectrum \cite{Komatsu:2001rj}. We use WMAP7 parameters: $\Omega_b=0.0449$, $\Omega_c = 0.222$, $h=0.71$, $\tau_{\rm re} = 0.088$, $A_s = 2.43 \cdot 10^{-9}$ and $n_s=0.963$. Since we want to concentrate on the signal generated at recombination, we do not consider second-order late-time effects such as vector, tensor and RS, that could correlate with the late ISW. 
As an illustration, in Fig.~\ref{fig:3dplot} we plot $b_{l_1l_2l_3}/(C_{l_1} C_{l_2} C_{l_3})^{1/2}$ with $l_3 = 1720$ as a function of $l_1$ and $l_2$.

To study the observability of the signal, we consider a full-sky cosmic variance limited experiment between $2$ and $l_{\rm max}$, without accounting for lensing. 
The Fisher matrix between two bispectra $i$ and $j$ is given by \cite{Komatsu:2001rj}
\be
F_{ij}  \equiv \sum_{2 \le l_1 \le l_2 \le l_3 \le l_{\rm max}} \frac{B^{(i)}_{l_1l_2l_3} B^{(j)}_{l_1l_2l_3}}{C_{l_1} C_{l_2} C_{l_3} \Delta_{l_1 l_2 l_3}}\;,
\ee
with
$\Delta_{l_1 l_2 l_3}=1,2,6$ for triangles with no, two or three equal sides. We find $F_{\rm  rec,rec}= 0.225 \ (0.504)$, $F_{\rm  rec,loc}= 0.038 \ (0.088)$ and $F_{\rm  loc,loc}= 0.046 \ (0.069)$ for $l_{\rm max} = 2000 \ (2500)$. 
The signal-to-noise ratio is given by 
\be
S/N \equiv F^{1/2}_{\rm  rec,rec} = 0.47 \ (0.71)\;;
\ee
most of it is concentrated in the squeezed configuration so that eq.~\eqref{rescaling} and the local correction \eqref{loc_red}  capture most of the signal.\footnote{The value of ``effective'' $f_{\rm NL}$ is given by $(F_{\rm rec, rec}/F_{\rm loc,loc})^{1/2}$. For $l_{\rm min} =100$ we find $f_{\rm NL}^{\rm eff} =  -3.82$, which agrees with \cite{Senatore:2008wk}. However, when $l_{\rm min} = 2$ it is smaller, $f_{\rm NL}^{\rm eff} =  -2.76$.}
Indeed, the contamination to a local signal agrees with \cite{Creminelli:2011sq,Bartolo:2011wb},\footnote{The local bias computed using eqs.~\eqref{loc_red} and \eqref{rescaling} with  $l_1 \le 60$ and $10 l_1 \le l_{2,3}$ is $f_{\rm NL}^{\rm loc} = 0.77 \ (1.16)$.}
\be
f_{\rm NL}^{\rm loc} \equiv F_{\rm  loc,rec}/F_{\rm  loc,loc} = 0.82 \ (1.27)\;,
\ee
and will negligibly  contaminate Planck's searches for local non-Gaussianity.

\vskip.1cm
{\em \bf Discussion:} We have presented the most complete calculation of second-order effects at recombination. It confirms that the projection on the local shape from recombination is well described by using the analytic estimate given in \cite{Creminelli:2011sq,Bartolo:2011wb,Lewis:2012tc} and suggests that claims of a large local contamination may be biased by numerical errors. Our calculation did not include lensing and other late-time integrated effects which correlate with the linear ISW effect. Since, for instance, lensing at recombination partially cancels with the rescaling of the coordinates \cite{Creminelli:2004pv,Creminelli:2011sq}, we must include all these effects to correctly estimate the total bias and the observability of the signal. We have shown that \CL~is a roboust tool to do this and, more generally, study second-order perturbations in cosmology.

\vskip.1cm
\emph{Acknowledgements:} We wish to thank F.~Bernardeau, P.~Creminelli, R.~Crittenden, C.~Fidler, A.~Lewis, G.~Pettinari and L.~Senatore for useful discussions and suggestions; we are particularly in debt to C. Pitrou for his help and advice on the code CMBquick. ZH is funded by the ANR {\em Chaire d'excellence Junior} ``CMBsecond''.

%%%%%%%%%%%%%%%%%%%%%%%%%%%%%%%%%%%%%%%%%%%%%%%%%%%%%%%%%%%%%%%%%%%%%
%%%% Bibliography

%\newpage

\end{document}